\documentclass[aip,jap,preprint]{revtex4-1}

\usepackage{graphicx}
\usepackage{dcolumn}
\usepackage{bm,bbm}

\begin{document}

\title{All-optical $\mathcal{PT}$-symmetric amplitude to phase modulator} 

\author{Oscar Ignacio Zaragoza Guti\'errez}
\affiliation{ Departamento de F\'isica, Universidad de Guadalajara, Revoluci\'on 1500, Guadalajara, Jal. CP 44420, M\'exico}
\altaffiliation{Verano de la Investigaci\'on Cient\'ifica, Academia Mexicana de Ciencias.}

\author{Luis Felipe Salinas Mendoza }
\affiliation{ Departamento de Ingenier\'ia Electr\'onica, Instituto Tecnol\'ogico de L\'azaro C\'ardenas, Melchor Ocampo 2555, L\'azaro C\'ardenas, Mich. CP 60950,  M\'exico }
\altaffiliation{Verano de la Investigaci\'on Cient\'ifica y Tecnol\'ogica del Pac\'ifico.}

\author{B. M. Rodr\'iguez-Lara}
\affiliation{Instituto Nacional de Astrof\'{\i}sica, \'Optica y Electr\'onica, Calle Luis Enrique Erro No. 1, Sta. Ma. Tonantzintla, Pue. CP 72840, M\'exico}
\email{bmlara@inaoep.mx}

\date{\today}

\begin{abstract}
We study electromagnetic field propagation through a planar three-waveguide coupler with linear gain and loss, in a configuration that is the optical analog of a quantum  $\mathcal{PT}$-symmetric system, and provide its closed-form analytic propagator. 
At an specific propagation length, we show that the device provides all-optical amplitude to phase modulation with a $\pi$ modulation range, if an extra binary phase is allowed in the reference signal, as well as phase to amplitude modulation, with an amplitude modulation range that depends linearly on the gain-to-coupling ratio of the system.
\end{abstract}

%\pacs{}

\maketitle 

% % % % % % % % % % % % % % % % % % % % % % % % % % % % % % % % % % % % % % % % % % % % % % % % % % % %
% Introduction
% % % % % % % % % % % % % % % % % % % % % % % % % % % % % % % % % % % % % % % % % % % % % % % % % % % %
%\section{Introduction}

Optical communications are a necessity of modern life.\cite{Liu2010p113,Benner2010} 
All-optical systems based on monolithic photonic integrated circuits may prove a compact, robust and ultrafast alternative with low-dissipation to some optoelectronic devices.\cite{Yu2015p031102}
For example, since the report of the optical directional couplers with attenuated waveguides\cite{Somekh1973p46} and the nonlinear coherent coupler,\cite{Jensen1982p1580} three-waveguide nonlinear directional couplers, based on the Kerr nonlinearity, have been shown to provide all-optical spatial switching\cite{Finlayson1990p2276,Stegeman1990p95,Artigas1996p53,Chen1997p287,Liu2003p2930,Khan2008p9417,Tao2011p071104} and logic gates.\cite{Menezes2007p1191,Menezes2007p107,Coelho2013p731}

Here, we are interested in the planar three-waveguide coupler with a gain mechanism different from the Kerr nonlinearity. 
In particular, we are going to study waveguides with complex effective refractive index, in other words, complex effective first-order susceptibilities, such that we have effective linear gain or loss in each waveguide. 
Such a system may be described by coupled-mode theory\cite{Somekh1973p46,Agarwal2012p031802,RodriguezLara2014p013802,RodriguezLara2015p068014} in the form of the differential equation set,
\begin{eqnarray}
- i \frac{d}{dz}\mathcal{E}_{0}(z) &=& n_{0} \mathcal{E}_{0}(z) + g \mathcal{E}_{1}(z),  \\
- i \frac{d}{dz}\mathcal{E}_{1}(z) &=& n_{1} \mathcal{E}_{1}(z) + g \left[ \mathcal{E}_{0}(z) + \mathcal{E}_{2}(z) \right], \\
- i \frac{d}{dz}\mathcal{E}_{2}(z) &=& n_{2} \mathcal{E}_{2}(z) + g \mathcal{E}_{1}(z), 
\end{eqnarray}
where the complex field amplitude at the $j$-th waveguide is given by $\mathcal{E}_{j}(z)$, the constant effective refractive index is $n_{j} = \omega_{j} + i \gamma_{j}$, with real parameters $\omega_{j}$ and $\gamma_{j}$, and the constant effective coupling between waveguide modes is $g$.
While it is possible to deal with this coupler, we will focus on the so-called $\mathcal{PT}$-symmetric optical systems.\cite{ElGanainy2007p2632,Guo2009p093902,Longhi2009p123601,Ruter2010p192,ElGanainy2013p161105,Longhi2014p1697}
In our device, that translates into effective refractive indices of the extremal waveguides that are complex conjugates of each other and that of the central waveguide is real. 
This model can be realized by engineering a three-waveguide coupler where the real part of the effective refractive index is $\omega_{0} = \omega_{2} = \omega$ and the gain-loss at each waveguide fulfills $\gamma_{1} = \left( \gamma_{0} + \gamma_{2} \right)/2$ and $\gamma =  \left( \gamma_{0} - \gamma_{2} \right)/2$, such that a solution of the form,
\begin{eqnarray}
\mathcal{E}_{j}(z) = e^{ i \left(  \omega_{1} + i \gamma_{1} \right)  z} E_{j}(z),
\end{eqnarray}
delivers an effective coupled-mode differential set, 
\begin{eqnarray}
- i \frac{d}{dz} E_{0}(z) &=& \left( \omega - \omega_{1} + i \gamma \right) E_{0}(z) + g E_{1}(z),  \\
- i \frac{d}{dz} E_{1}(z) &=&  g \left[ E_{0}(z) + E_{2}(z) \right], \\
- i \frac{d}{dz} E_{2}(z) &=& \left( \omega - \omega_{1} - i \gamma \right) E_{2}(z) + g E_{1}(z).
\end{eqnarray}

Amplitude to phase modulation is another key element in communications and, to the best of our knowledge, an all-optical solution involving $\mathcal{PT}$-symmetric three-waveguide couplers has not been reported in the literature.
For that reason, in the following, we will discuss propagation through the three-waveguide $\mathcal{PT}$-symmetric coupler in the case of waveguides with identical real part of the effective refractive index, provide a closed-form analytic propagator for the device and explore its use as an amplitude to phase and phase to amplitude modulator.

% % % % % % % % % % % % % % % % % % % % % % % % % % % % % % % % % % % % % % % % % % % % % % % % % % % %
% Solution
% % % % % % % % % % % % % % % % % % % % % % % % % % % % % % % % % % % % % % % % % % % % % % % % % % % %
%\section{Field propagation}

For the sake of simplicity, we will consider waveguides with identical real part of the effective refractive index, $\omega_{1} = \omega$, such that we obtain the coupled-mode set,
\begin{eqnarray}
-i \frac{d}{d\zeta} \bm{E}(\zeta) =  \mathbbm{H} \bm{E}(\zeta),
\end{eqnarray}
where we have defined a scaled propagation distance, $\zeta = \lambda z$, in order to have a single control parameter in the form of the ratio between the imaginary part of the effective refractive index and the coupling parameter, $\xi = \gamma / \lambda$, henceforth named gain-to-coupling ratio, and used the shorthand notation, 
\begin{eqnarray}
 \mathbbm{H} = \left(
\begin{array}{ccc}
i \xi & 1 & 0 \\
1 & 0 & 1 \\
0 & 1 & - i \xi
\end{array}
\right), \quad  \bm{E}(\zeta)  = \left( \begin{array}{c} E_{0}(\zeta),\\ E_{1}(\zeta),\\ E_{2}(\zeta) \end{array}\right),
\end{eqnarray} 
for the mode coupling matrix and the complex field amplitudes vector, in that order.
It is straightforward to show that this mode coupling matrix has two nonzero eigenvalues, $ \pm \Omega$, with
\begin{eqnarray}
\Omega = \sqrt{ 2 - \xi^{2}}.
\end{eqnarray}
The eigenvalues will be real when the gain to coupling ratio fulfills $\xi < \sqrt{2}$, completely degenerate for $\xi=\sqrt{2}$, and complex if  $\xi > \sqrt{2}$.
In other words, $\mathcal{PT}$-symmetry is broken in the latter case.

The mode coupling matrix, $\mathbbm{H}$, does not depend on the scaled propagation distance, $\zeta$, thus, we can calculate the propagated complex field amplitudes as,\cite{RodriguezLara2015p068014} 
\begin{eqnarray}
\bm{E}(\zeta)  = \mathbbm{U}(\zeta) \bm{E}(0), \quad  \mathbbm{U}(\zeta)= e^{i  \mathbbm{H} \zeta},
\end{eqnarray}
where the input complex field amplitudes are collected in the vector $ \bm{E}(0)$.
The propagator, $ \mathbbm{U}(\zeta)$, can be calculated from its Taylor series,
\begin{eqnarray}
 \mathbbm{U}(\zeta) &=& \sum_{j=0}^{\infty} \frac{(i \zeta)^{j}}{j!}  \mathbbm{H}^{j},\\
&=&  \left\{  \begin{array}{ll}
\mathbbm{1} +  \frac{i}{\Omega} \sin \Omega  \zeta ~ \mathbbm{H} + \frac{1}{\Omega^2} \left( \cos \Omega  \zeta - 1 \right)  \mathbbm{H}^{2}, & \xi \neq \sqrt{2}, \\
\mathbbm{1} + i  \zeta \mathbbm{H} - \frac{1}{2} \zeta^2 \mathbbm{H}^2, & \xi = \sqrt{2} \end{array}, \right.
\end{eqnarray} 
where  the identity matrix is given by $\mathbbm{1}$ and  we used the Cayley-Hamilton theorem,\cite{RodriguezLara2005p87}  a square matrix satisfies its own characteristic equation, in this particular case,
\begin{eqnarray}
 \mathbbm{H}^{3} = \Omega^{2}  \mathbbm{H}.
\end{eqnarray}

\begin{figure}
\includegraphics[scale=1]{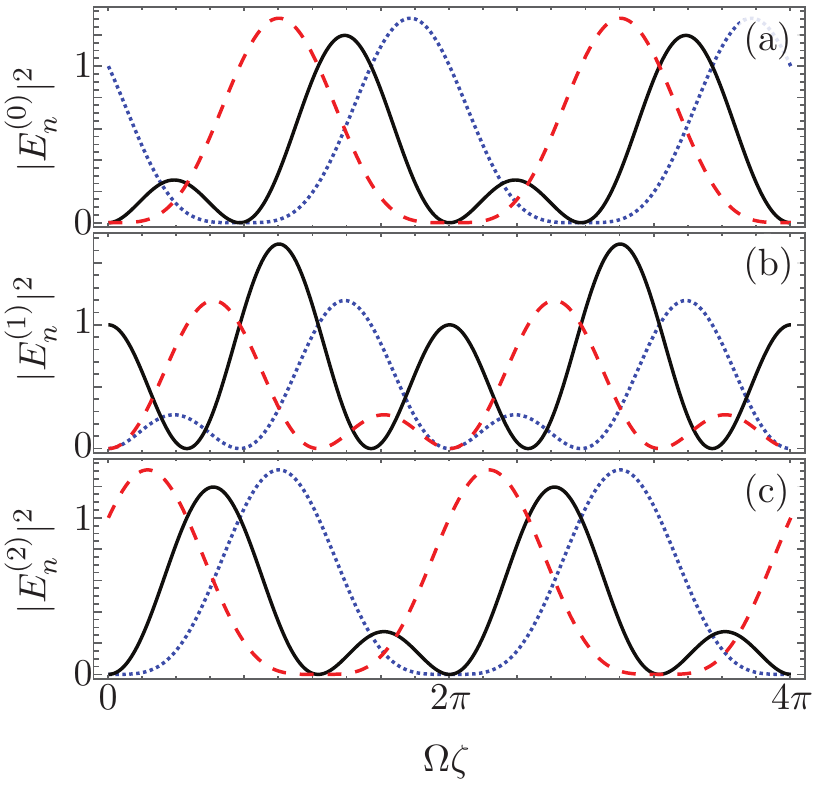}
\caption{(Color online) Squared response to impulse function, $\vert E_{n}^{(m)} \vert^2$, at the zeroth (dotted blue), $n=0$, first (solid black), $n=1$, and second (dashed red), $n=2$, waveguides for an initial field impinging at the (a) zeroth, $m=0$, (b) first, $m=1$, and (c) second, $m=2$, waveguide of a coupler with gain-to-coupling ratio $\xi =0.5$ that keeps $\mathcal{PT}$-symmetry. \label{fig:Fig1}}
\end{figure}

After some algebra, it is possible to provide the propagation as a response to an impulse function, that is, the complex field amplitude at the $n$-th waveguide given that the input complex field amplitude impinged at just the $m$-th waveguide, 
\begin{eqnarray}
E_{n}^{(m)}(\zeta) &=& \mathbbm{U}_{n,m}(\zeta), \quad n, m = 0,1,2,\\
&=& \frac{\alpha^{m+n}}{\alpha^{0}}\sqrt{ \left( \begin{array}{cc} 2 \\ m \end{array} \right) \left( \begin{array}{cc} 2 \\ n \end{array} \right)} ~_{2}F_{1}\left[-m,-n,-2,-\frac{\alpha_{0}}{\alpha^{2}}\right],
\end{eqnarray}
with Gauss hypergeometric function, $_2F_{1}(a,b,c,d)$,\cite{Lebedev1965} and auxiliary functions,
\begin{eqnarray}
\alpha_{0} &=& \left\{  \begin{array}{ll}
\Omega^2/\beta^{2}, & \xi \neq \sqrt{2}, \\
2/\left( \sqrt{2} - \zeta \right)^{2}, & \xi = \sqrt{2},  \end{array}\right.  \\
\alpha &=& \left\{  \begin{array}{ll}
i \sqrt{2} / \beta, & \xi \neq \sqrt{2}, \\
i \zeta/\left(\sqrt{2} - \zeta \right), & \xi=\sqrt{2}, \end{array}\right.  \\
\beta &=& \Omega \cos \frac{1}{2} \Omega \zeta  - \xi \sin \frac{1}{2} \Omega \zeta.
\end{eqnarray}
Thus, even with amplification and losses, we expect periodic oscillations of the complex field amplitudes in the $\mathcal{PT}$-symmetric regime, $\xi < \sqrt{2}$ with a period equal to $2 \pi \Omega^{-1}$, Fig. \ref{fig:Fig1}, that will disappear as the $\mathcal{PT}$-symmetry is broken, $\xi \ge \sqrt{2}$, and  gain dominates, Fig. \ref{fig:Fig1}.
We conducted both a numerical solution of the differential equation set and propagation with the analytic results and obtained good agreement between both methods.

\begin{figure}
\includegraphics[scale=1]{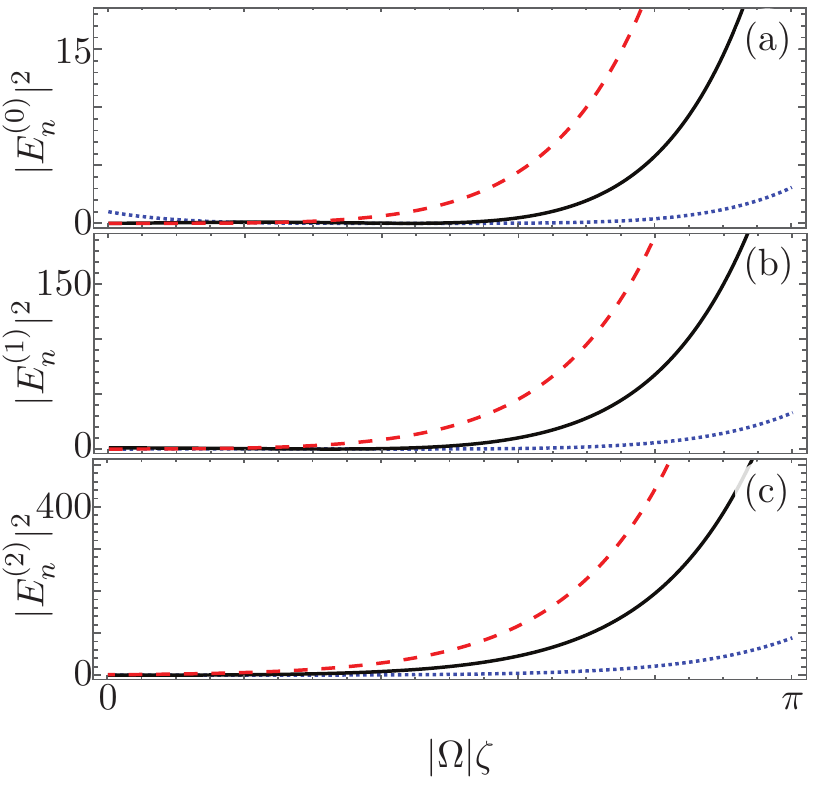}
\caption{(Color online) Same as Fig. \ref{fig:Fig2} but for a gain-to-coupling ratio $\xi = 1.25 \sqrt{2}$ that breaks $\mathcal{PT}$-symmetry. \label{fig:Fig2}}
\end{figure}

% % % % % % % % % % % % % % % % % % % % % % % % % % % % % % % % % % % % % % % % % % % % % % % % % % % %
% Applications
% % % % % % % % % % % % % % % % % % % % % % % % % % % % % % % % % % % % % % % % % % % % % % % % % % % %
%\section{$\mathcal{PT}$-symmetric modulator}

In the following, we will focus on just the $\mathcal{PT}$-symmetric regime, $0 \le \xi < \sqrt{2}$, because the device always behaves like a directional amplifier in the broken symmetry phase, Fig. \ref{fig:Fig2}. 
We want to bring forward that the response to impulse function, Fig. \ref{fig:Fig1}, reveals that, at a certain scaled distance, 
\begin{eqnarray}
 \zeta_{f} =  \frac{ \arccos \left(\xi^{2} - 1 \right) + 2 n \pi}{\Omega}, \quad n=0,1,2, \ldots,
\end{eqnarray}
the output complex field amplitudes are quite simple,
\begin{eqnarray}
E_{0}(\zeta_{f}) &=& - E_{2}(0) , \\
E_{1}(\zeta_{f}) &=& - E_{1}(0) + i 2 \xi E_{2}(0), \\
E_{2}(\zeta_{f}) &=& - E_{0}(0) +  2 \xi \left[ i E_{1}(0) + \xi E_{2}(0) \right]. 
\end{eqnarray} 
If we consider initial complex field amplitudes,
\begin{eqnarray}
E_{0}(0) = \mathcal{A}_{0} e^{i \phi_{1}}, ~
E_{1}(0) = \mathcal{A}_{1} e^{i \phi_{1}}, ~
E_{2}(0) = 0,
\end{eqnarray}
with real amplitudes and phases, $\mathcal{A}_{j} \ge 0$ and $0 \le \phi_{j} < 2 \pi$, the fields at the output are given by,
\begin{eqnarray}
E_{0}(\zeta_{f}) &=& 0 , \\
E_{1}(\zeta_{f}) &=& \mathcal{A}_{1} e^{i \left( \phi_{1} + \pi \right)}, \\
E_{2}(\zeta_{f}) &=& \mathcal{A}_{0} e^{i \left( \phi_{0} + \pi \right)} +  2 \xi \mathcal{A}_{1} e^{i \left( \phi_{1} + \frac{\pi}{2} \right)}.
\end{eqnarray} 
Thus, it is possible to convert amplitude and phase differences between the signal, $E_{0}(0)$, and reference, $E_{1}(0)$, input fields to phase and amplitude differences at the output field, 
\begin{eqnarray}
\vert E_{2}(\zeta_{f}) \vert  &=& \sqrt{ \mathcal{A}_{0}^{2} +  4 \xi^{2}  \mathcal{A}_{1}^{2} - 4 \xi \mathcal{A}_{0} \mathcal{A}_{1} \sin \delta }, \\
\arg \left[ E_{2}(\zeta_{f}) \right] &=& \arctan \frac{\mathcal{A}_{0} \sin \phi_{0} - 2 \xi \mathcal{A}_{1} \cos \phi_{1}}{\mathcal{A}_{0} \cos \phi_{0} + 2 \xi \mathcal{A}_{1} \sin \phi_{1}},
\end{eqnarray}
with the phase difference between signal and reference given by $\delta= \phi_{0} - \phi_{1}$.
We will take the argument function as modulo $2 \pi$ to have a phase in the range $[0,2 \pi)$.
Note, we will assign a null phase value to zero fields.

%\subsection{Amplitude to phase conversion}

\begin{figure}
\includegraphics[scale=1]{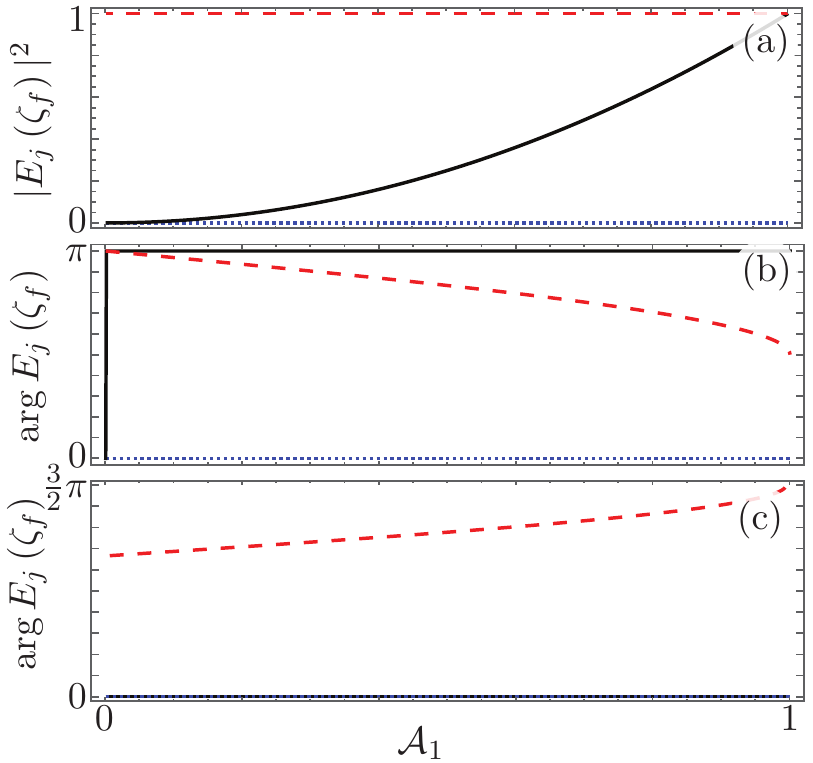}
\caption{(Color online) (a) Squared field amplitude, $\vert E_{j}(\zeta_{f}) \vert^2$,  and (b-c) phase, $\arg E_{j}(\zeta_{f})$,  with $j=0$ (dotted blue), $j=1$ (solid black), and $j=2$ (dashed red), as a function of the reference field (a) $E_{1}(0)= \mathcal{A}_{1}$  and (b) $E_{1}(0)=- \mathcal{A}_{1}$  for a $\mathcal{PT}$-symmetric amplitude to phase modulator with gain-to-coupling parameter $\xi = 0.5$. \label{fig:Fig3}}
\end{figure}

Let us consider first the case of input fields of proportional amplitudes,
\begin{eqnarray}
E_{0}(0) = \sqrt{1 - \mathcal{A}_{1}^2}, ~
E_{1}(0) = \pm \mathcal{A}_{1}, ~
E_{2}(0) = 0.
\end{eqnarray}
Note that we have added a binary phase on the reference field, $E_{1}(0)$, that will help us in the future.
The output field amplitude and phase are given by the following, 
\begin{eqnarray}
\vert E_{2}(\zeta_{f}) \vert  &=& \sqrt{1 - \left( 4 \xi^{2} - 1 \right) \mathcal{A}_{1}^2}, \\
\arg \left[ E_{2}(\zeta_{f}) \right] &=& \arctan \mp 2 \xi \eta,
\end{eqnarray}
where the phase is a function of the reference-to-signal amplitude ratio, \begin{eqnarray}
 \eta = \frac{\mathcal{A}_{1}}{\sqrt{1- \mathcal{A}_{1}^{2}}}.
\end{eqnarray}
For any given gain-to-coupling ratio, each of the two binary phase options provides a $\pi/2$ modulation range but in different phase regions,
\begin{eqnarray}
\mathcal{V}_{\phi} &=& \left[ \arg E_{2}(\zeta_{f}) \right]_{\max} - \left[ \arg E_{2}(\zeta_{f}) \right]_{\min}, \\
&=& \left\{  \begin{array}{ll} \pi - \pi/2, & E_{1}(0) = \mathcal{A}_{1},  \\
3 \pi /2 - \pi, & E_{1}(0) = - \mathcal{A}_{1},    \end{array} \right.  \\
&=& \frac{\pi}{2},
\end{eqnarray}
Thus, the introduction of the binary phase allows for continuous modulation of phase in a $\pi$ range.
Figure \ref{fig:Fig3} shows an example of amplitude to phase modulation with a gain-to-coupling ratio $\xi=0.5$ that provides an stable output field amplitude, Fig. \ref{fig:Fig3}(a), for all the phase modulation range, Fig. \ref{fig:Fig3}(b) and Fig. \ref{fig:Fig3}(c).

%\subsection{Phase to amplitude conversion}

\begin{figure}
\includegraphics[scale=1]{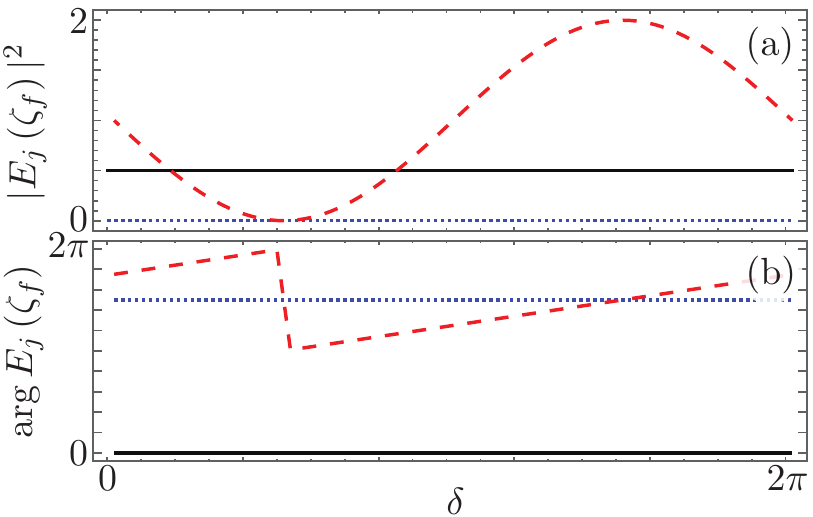}
\caption{(Color online) (a) Squared field amplitude, $\vert E_{j}(\zeta_{f}) \vert^2$,  and (b) phase, $\arg E_{j}(\zeta_{f})$,  with $j=0$ (dotted blue), $j=1$ (solid black), and $j=2$ (dashed red), as a function of the initial phase difference, $\delta$, between signal, $E_{0}(0) = e^{i \delta}/ \sqrt{2}$, and  reference,  $E_{1}(0) = 1/ \sqrt{2}$, fields for a $\mathcal{PT}$-symmetric phase to amplitude modulator with gain-to-coupling parameter $\xi = 0.5$. \label{fig:Fig4}}
\end{figure}

The converse situation is provided when the input fields have the same amplitude but different phase, 
\begin{eqnarray}
E_{0}(0) = e^{i \delta} \mathcal{A}_{1},  ~
E_{1}(0) = \mathcal{A}_{1}, ~
E_{2}(0) = 0.
\end{eqnarray}
Here the output field amplitude and phase are a function of the phase difference, 
\begin{eqnarray}
\vert E_{2}(\zeta_{f}) \vert  &=& \sqrt{1+ 4 \xi \left(\xi-1\right)\sin \delta} ~\mathcal{A}, \\
\arg \left[ E_{2}(\zeta_{f}) \right] &=& \arctan \frac{2 \xi - \sin \delta }{\cos \delta}.
\end{eqnarray}
Thus, we are mapping the phase difference to both the amplitude and phase of the output field.
The range of the phase to amplitude modulation can be measured by an equivalent of the interferometric visibility,
\begin{eqnarray}
\mathcal{V}_{a} &=& \frac{\vert E_{2}(\zeta_{f}) \vert^2_{\max} - \vert E_{2}(\zeta_{f}) \vert^2_{\min}}{\vert E_{2}(\zeta_{f}) \vert^2_{\max} + \vert E_{2}(\zeta_{f}) \vert^2_{\min}}, \\
&=& \frac{8\xi \mathcal{A}_{0} \mathcal{A}_{1}}{ \mathcal{A}_{0}^{2} +  4 \xi^{2}  \mathcal{A}_{1}^{2}},
\end{eqnarray}
where we can see that the gain-to-coupling ratio plays a fundamental role. 
Figure \ref{fig:Fig4} shows the amplitude and phase of the output fields  as a function of the initial phase difference between signal and reference fields, for a gain-to-coupling ration $\xi = 0.5$ with equal initial complex field amplitudes, $\mathcal{A}_{1} = \mathcal{A}_{2} = 1/\sqrt{2}$.

% % % % % % % % % % % % % % % % % % % % % % % % % % % % % % % % % % % % % % % % % % % % % % % % % % % %
% Conclusion
% % % % % % % % % % % % % % % % % % % % % % % % % % % % % % % % % % % % % % % % % % % % % % % % % % % %
%\section{Conclusion}

In summary, we have studied the propagation of electromagnetic fields through three coupled waveguides with gain. 
The real and imaginary parts of the couled mode effective refractive index of the waveguides are chosen to produce an optical analog of quantum $\mathcal{PT}$-symmetry.
We focus in the particular case of refractive indices with identical real parts and provide a closed form analytic propagator for the device.
Then, we show that the device can be used as an optical oscillator or directional amplifier inside and outside the $\mathcal{PT}$-symmetric regime, in that order.

Furthermore, we show that there exists a characteristic length that delivers a two-input, single-output, $\mathcal{PT}$-symmetric device providing all-optical amplitude (phase) to phase (amplitude) modulation.
The amplitude to phase configuration provides a phase modulation range of $\pi/2$ that can be increased to $\pi$ if a binary phase is introduced into the reference signal.
The phase to amplitude mode provides an amplitude modulation range that depends linearly on the gain-to-coupling ratio of the waveguide array.

% % % % % % % % % % % % % % % % % % % % % % % % % % % % % % % % % % % % % % % % % % % % % % % % % % % %
% Acknowledgements
% % % % % % % % % % % % % % % % % % % % % % % % % % % % % % % % % % % % % % % % % % % % % % % % % % % %
%\begin{acknowledgments}
OIZG acknowledges financial support from the Academia Mexicana de Ciencias through the Verano de la Investigaci\'on Cient\'ifica program and LFSM from the Instituto Tecnol\'ogico de L\'azaro C\'ardenas through the Verano de la Investigaci\'on Cient\'ifica y Tecnol\'ogica del Pac\'ifico program. BMRL acknowledges fruitful discussion with J. David S\'anchez de la Llave.
%\end{acknowledgments}

% % % % % % % % % % % % % % % % % % % % % % % % % % % % % % % % % % % % % % % % % % % % % % % % % % % %
% Bibliography
% % % % % % % % % % % % % % % % % % % % % % % % % % % % % % % % % % % % % % % % % % % % % % % % % % % %
%\bibliography{D:/ExternalHD/Bibliography/references}

%

\end{document}